\documentclass{article}

\usepackage{epsfig}
\usepackage{graphicx}
\usepackage{amssymb}
\usepackage{url}
\usepackage{amsfonts,amsbsy}
\usepackage{psfrag}
\usepackage{amsmath}

\def\e{\begin{equation}} 
\def\f{\end{equation}} 
\def\ea{\begin{eqnarray}} 
\def\fa{\end{eqnarray}} 

\def\##1{{\mbox{\textbf{#1}}}}
\def\%#1{{\mbox{\boldmath $#1$}}}

\def\=#1{{\overline{\overline{\mathsf #1}}}}

\def\/{\over}
\def\*{^{\displaystyle*}}

\def\.{\cdot}
\def\x{\times}
\def\:{\over}

\def\D{\nabla}

\def\ra{\rightarrow}

\def\Ra{\Rightarrow}

\def\l#1{\label{eq:#1}}
\def\r#1{(\ref{eq:#1})}
\def\am{\left(\begin{array}{c}}
\def\amm{\left(\begin{array}{cc}}
\def\ammm{\left(\begin{array}{ccc}}
\def\ammmm{\left(\begin{array}{cccc}}
\def\a{\end{array}\right)}

\def\add{\left|\begin{array}{cc}}
\def\addd{\left|\begin{array}{ccc}}
\def\adddd{\left|\begin{array}{cccc}}
\def\ad{\end{array}\right|}

\def\A{\alpha}
\def\B{\beta}
\def\de{\delta}

\def\E{\epsilon}
\def\g{\gamma}

\def\h{\eta}

\def\la{\lambda}

\def\M{\mu}
\def\o{\omega}

\def\VR{\varrho}

\def\�#1{\underline{\bf #1\mit}}

\begin{document}

\title{Electromagnetic Boundaries\\ with PEC/PMC Equivalence}
\author{Ismo V.~Lindell and Ari~Sihvola\\ {\tt ismo.lindell@aalto.fi\quad ari.sihvola@aalto.fi}\\
{\it Department of Radio Science and Engineering}\\ {\it Aalto University, School of Electrical Engineering}\\
{\it Espoo, Finland}}

\maketitle
\begin{abstract}
The most general electromagnetic boundary, defined by linear and  local boundary conditions, is defined in terms of conditions which can be called generalized impedance boundary conditions. Requiring that the boundary be equivalent to PEC and PMC boundaries for its two eigen-plane waves, which property is known to exist for many of its special cases, it is shown that the recently introduced Generalized Soft-and-Hard/DB (GSHDB) boundary is the most general boundary satisfying this property.
\end{abstract}

\section{Introduction}

Boundary surface is a conceptual two-dimensional structure in which electromagnetic sources, induced by the external field, are related by some intrinsic mechanism. As sources we may assume electric and magnetic surface currents, $\#J_{es}, \#J_{ms}$, and electric and magnetic surface charges, $\VR_{es}, \VR_{ms}$. 
When the unit vector normal to the boundary surface is denoted by $\#e_3$, the fields outside the boundary are related to the surface sources by the conditions \cite{EWT}
\e \#e_3\x\#E = -\#J_{ms},\ \ \ \ \#e_3\x\#H = \#J_{es}, \l{e3E}\f
\e \#e_3\.\#D = \VR_{es},\ \ \ \  \#e_3\.\#B=\VR_{ms}. \l{e3D}\f
For simplicity we assume a planar boundary and constant unit vectors $\#e_1,\#e_2,\#e_3$ making an orthonormal basis. Assuming time-harmonic fields with time dependence $\exp(j\o t)$, the sources obey the continuity conditions
\e \D\.\#J_{es} = -j\o\VR_{es},\ \ \ \ \D\.\#J_{ms}=-j\o\VR_{ms}, \f
following from the Maxwell equations and \r{e3E}, \r{e3D}.

Let us assume that the relations between the source quantities, set by the boundary structure, are linear and local and can be expressed by linear algebraic equations. Because of the relations \r{e3E} and \r{e3D}, the fields at the boundary are related in a certain manner forming the boundary conditions which are linear and local. Considering the basic problem of a field incident to the boundary,  due to the Huygens principle, the reflected field is uniquely determined when two scalar components of the field vectors tangential to the surface are known. Thus, the boundary conditions must be of the form of two scalar conditions between the fields at the surface. Under the assumption of linearity and locality, the most general boundary conditions can be assumed to have the form 
\e \A\#e_3\.\#B+ \B\#e_3\.\#D+\#a_t\.\#E + \#b_t\.\#H =0,\l{ximp1}\f
\e \g\#e_3\.\#B+ \de\#e_3\.\#D+ \#c_t\.\#E+ \#d_t\.\#H=0, \l{ximp2}\f
relating the normal components of $\#D$ and $\#B$ vectors and tangential components of $\#E$ and $\#H$ vectors in terms of four vectors and four scalars. The vectors tangential to the boundary surface are denoted by the subscript $()_t$. The form of \r{ximp1} and \r{ximp2} can be simplified by eliminating $\#e_3\.\#B$ on one hand, and $\#e_3\.\#D$ on the other hand, whence the most general form can be written as
\e \A\#e_3\.\#B+ \#a_t\.\#E + \#b_t\.\#H =0,\l{ximpB}\f
\e \de\#e_3\.\#D+ \#c_t\.\#E+ \#d_t\.\#H=0,. \l{ximpD}\f

\section{Boundary conditions}

Let us consider the boundary conditions \r{ximpB} and \r{ximpD} for some special choices of the two scalars and four tangential vectors.

\begin{itemize}
\item The perfect electric conductor (PEC) conditions,  $\A=\de=0$, $\#b_t=\#d_t=0$, $\#a_t=\#e_1$, $\#c_t=\#e_2$,
\e \#e_1\.\#E=0,\ \ \#e_2\.\#E=0,\ \ \ \ \Ra\ \ \ \ \#e_3\x\#E=0. \f
\item The perfect magnetic conductor (PMC) conditions, $\A=\de=0$, $\#a_t=\#c_t=0$, $\#b_t=\#e_1$, $\#d_t=\#e_2$,
\e \#e_1\.\#H=0,\ \ \#e_2\.\#H=0,\ \ \ \ \Ra\ \ \ \ \#e_3\x\#H=0. \f
\item The perfect electromagnetic conductor (PEMC) conditions \cite{PEMC}, $\A=\de=0$, $\#a_t=M\#b_t=M\#e_1$, $\#c_t=M\#d_t=M\#e_2$,
\e \#e_1\.\#(\#H+M\#E)=0,\ \ \#e_2\.\#(\#H+M\#E)=0,\ \ \ \ \Ra\ \ \ \ \#e_3\x(\#H+M\#E)=0. \f
\item The DB conditions \cite{DB},  $\A=\de=1$, $\#a_t=\#b_t=\#c_t=\#d_t=0$,
\e \#e_3\.\#D=0,\ \ \ \ \#e_3\.\#B=0. \f
\item The soft-and-hard (SH) conditions \cite{SHS1},  $\A=\de=0$, $\#a_t=\#d_t=\#e_1$, $\#b_t=\#c_t=0$,
\e \#e_1\.\#E=0,\ \ \ \ \#e_1\.\#H=0. \f
\item The generalized soft-and-hard (GSH) conditions \cite{GSH},  $\A=\de=0$, $\#b_t=\#c_t=0$,
\e \#a_t\.\#E=0,\ \ \ \ \#d_t\.\#H=0. \l{GSH}\f
\item The soft-and-hard/DB (SHDB) conditions \cite{SHDB}, $\A=\de$, $\B=\g=0$, $\#b_t=\#c_t=0$, $\#a_t=-\#d_t$,
\e \A\#e_3\.\#B + \#a_t\.\#E=0,\ \ \ \ \A\#e_3\.\#D - \#a_t\.\#H=0. \l{SHDB}\f
\item The generalized soft-and-hard/DB (GSHDB) conditions \cite{GSHDB},  $\#b_t=\#c_t=0$, 
\e \A\#e_3\.\#B + \#a_t\.\#E=0,\ \ \ \ \de\#e_3\.\#D +\#d_t\.\#H=0, \l{GSHDB}\f
\item The impedance conditions  $\A=\de=0$,
\e \#a_t\.\#E + \#b_t\.\#H =0,\ \ \ \ \#c_t\.\#E+ \#d_t\.\#H=0. \l{imp}\f
which can also be written as $\#e_3\x\#E = \=Z_s\.\#H$, with $\=Z_s=(\#a_t\#d_t-\#c_t\#b_t)/(\#e_3\.\#a_t\x\#c_t)$.
\end{itemize}

Comparing with \r{imp}, the form \r{ximpB}, \r{ximpD} can be called generalized impedance conditions. Because each tangential vector has two free parameters, the number of free parameters of the GSHDB boundary \r{GSHDB} is 4, for the impedance boundary \r{imp} it is 6 and for the generalized impedance boundary \r{ximpB}, \r{ximpD} it is 10.

One should note that non-local boundary conditions are not included in the definition \r{ximpB} and \r{ximpD}. For example the D'B' boundary defined by the conditions  \cite{D'B'}
\e \#e_3\.\D(\#e_3\.\#D)=0,\ \ \ \ \ \#e_3\.\D(\#e_3\.\#B)=0, \l{D'B'}\f
would require operator-valued scalars $\A$ and $\de$ in \r{ximpB} and \r{ximpD}.

\section{Plane-wave reflection}

Considering a time-harmonic plane wave incident to and reflecting from the boundary surface,
\e \#E^i(\#r)= \#E^i\exp(-j\#k^i\.\#r),\ \ \ \ \#E^r(\#r)= \#E^r\exp(-j\#k^r\.\#r), \f
with
\e \#k^i = \#k_t -k_3\#e_3,\ \ \ \ \#k^r=\#k_t + k_3\#e_3, \f
and applying the Maxwell equations, we can write the following relations for the total fields at the boundary surface,
\ea \o\#e_3\.\#B &=&\o\#e_3\.(\#B^i+\#B^r) = \#e_3\.\#k_t\x(\#E^i+\#E^r)= (\#e_3\x\#k_t)\.\#E, \\
 \o\#e_3\.\#D &=&\o\#e_3\.(\#D^i+\#D^r) = -\#e_3\.\#k_t\x(\#H^i+\#H^r)= -(\#e_3\x\#k_t)\.\#H. \fa
Substituting $\#e_3\.\#B$ and $\#e_3\.\#D$ in the generalized impedance conditions \r{ximpB} and \r{ximpD}, they obtain the form of the impedance conditions \r{imp} if, in the latter, we substitute
\e \#a_t\ra \#a_t + \A\#e_3\x\#k_t,\ \ \ \#d_t\ra\#d_t- \de\#e_3\x\#k_t. \l{gic}\f
Thus, if the vectors $\#a_t, \#d_t$ are allowed to have a similar linear dependence on the vector, $\#e_3\x\#k_t$, the impedance conditions \r{imp} represent the most general form of boundary conditions for a plane wave.

It has been previously shown that, for the generalized SHDB boundary \r{GSHDB}, there exist two eigenwaves, one of which is reflected from the boundary as from a PEC boundary and, the other one, as from a PMC boundary. This property is valid for all of its special cases, the SHDB boundary, the GSH boundary, the SH boundary and the DB boundary. Also, the PEMC boundary shares the same property whereas the most general impedance boundary doesn't. The property of PEC/PMC equivalence is most useful because, given any incident wave, the reflected wave can be easily found by decomposing the incident wave in its eigencomponents. The task is to find the restriction for the generalized impedance boundary which allows the boundary to be replaced by PEC and PMC boundaries for the respective eigenwaves. 

Invoking the results of \cite{GSHDB}, the relations between tangential components of the electric and magnetic fields of incident and reflected plane waves in an isotropic medium with parameters $\E_o,\M_o$ can be written as
\e \h_o\#H_t^i = -\=J_t\.\#E_t^i,\ \ \ \ \#E_t^i= \=J_t\.\h_o\#H_t^i, \l{Hi}\f
\e \h_o\#H_t^r = \=J_t\.\#E_t^r,\ \ \ \ \#E_t^r= -\=J_t\.\h_o\#H_t^r, \l{Hr}\f
where the dyadic $\=J_t$ is defined by
\e \=J_t = \frac{1}{k_ok_3}((\#e_3\x\#k_t)\#k_t + k_3^2\#e_3\x\=I_t).\f
For an eigenfield the tangential field components are multiples of one another. Defining $\#E^r_t=\la\#E^i_t$, from \r{Hi} and \r{Hr} we obtain $\#H_t^r=-\la\#H_t^i$. Thus, the PEC and PMC boundaries correspond to the respective eigenvalues $\la=-1$ and $\la=+1$. 

Let us first find under what restrictions to the four vectors $\#a_t, \#b_t, \#c_t$ and $\#d_t$ the eigenvalues corresponding to the impedance boundary \r{imp} are $+1$ and $-1$. Writing the conditions \r{imp} for the eigenfields as 
\e ((1+\la)\h_o\#a_t - (1-\la)\#b_t\.\=J_t)\.\#E_t^i = 0,\f
\e ((1+\la)\h_o\#c_t - (1-\la)\#d_t\.\=J_t)\.\#E_t^i =0, \f
to have solutions other than $\#E_t^i=0$, the bracketed vector expressions must be parallel. Thus, the eigenvalue $\la$ must satisfy the equation
\e \#e_3\.((1+\la)\h_o\#a_t - (1-\la)\#b_t\.\=J_t)\x((1+\la)\h_o\#c_t- (1-\la)\#d_t\.\=J_t)=0. \f
Let us require that it be satisfied for $\la=+1$ and $\la=-1$, which yields the two conditions:
\e \#e_3\.(\#a_t\x\#c_t)=0 \l{eac}\f
\e \#e_3\.((\#b_t\.\=J_t)\x(\#d_t\.\=J_t))=\#e_3\.((\#b_t\x\#d_t)\.\=J{}_t^{(2)})=\#e_3\.(\#b_t\x\#d_t)=0. \l{ebd}\f
In the latter equation we use the property $\=J_t^{(2)} = \#e_3\#e_3$ and rules of dyadic algebra \cite{Methods}.  \r{eac} and \r{ebd} show that, to obtain eigenvalues $\la=\pm1$, the tangential vectors $\#a_t$ and $\#c_t$ on one hand, and $\#b_t$ and $\#d_t$ on the other hand, must be linearly dependent, whence they must satisfy conditions of the form
\e A\#a_t+  C\#c_t=0,\ \ \ \ B\#b_t+ D\#d_t=0 \f
for some scalars $A - D$. Operating the impedance boundary conditions \r{imp} as
\e \amm A & C\\ B & D\a \amm \#a_t & \#b_t\\ \#c_t & \#d_t\a\.\am \#E\\ \#H\a = \am(A\#b_t+C\#d_t)\.\#H\\ (B\#a_t+D\#c_t)\.\#E\a = \am 0\\ 0\a, \l{cond}\f
the required boundary conditions must reduce to the form $\#a_t'\.\#E=0$ and $\#b_t'\.\#H=0$, which can be recognized as the generalized soft-and-hard (GSH) boundary conditions \r{GSH}. 

For the generalized impedance conditions \r{ximpB}, \r{ximpD}, we can make the substitutions \r{gic}, whence \r{cond} can be written as
\e \am(A\#b_t+C(\#d_t-\de\#e_3\x\#k_t))\.\#H\\ (B(\#a_t+\A\#e_3\x\#k_t)+D\#c_t)\.\#E\a = \am 0\\ 0\a. \f
Applying plane-wave relations, these conditions can be expressed as
\e \am(A\#b_t+C\#d_t)\.\#H +\o C\de\#e_3\.\#D\\ (B\#a_t+D\#c_t)\.\#E + \o B\A\#e_3\.\#B\a = \am 0\\ 0\a, \f
which have the form of the generalized soft-and-hard/DB conditions \r{GSHDB}.

\section{Conclusion}

The task taken in this paper was to find the most general linear and local boundary conditions which allow plane waves to be split in two components one of which is reflected as from the PEC boundary and, the other one, as from the PMC boundary. For this, the most general linear and local boundary conditions were first expressed in a form which can be called generalized impedance boundary conditions. Since PEC and PMC boundary conditions for a plane wave yield the reflection coefficients $\pm1$, the problem was reduced to finding out corresponding restrictions for the generalized impedance boundary. The outcome was that the generalized impedance conditions must actually be of the form of what have been called generalized soft-and-hard/DB conditions, previously studied by these authors. However, one should note that, because of the assumption of locality, there may exist other solutions as well. For example, the non-local D'B' boundary conditions \r{D'B'} are also known to share this PEC/PMC property \cite{D'B'}. While the result of this paper is mainly of theoretical interest, realizations of various boundary conditions as metasurfaces have been reported in \cite{Caloz} -- \cite{Frezza}, and applications have been pointed out in \cite{Kong08} -- \cite{Kildal09}.


\begin{thebibliography}{99}


\bibitem{EWT}  Kong, J. A., {\it Electromagnetic Wave Theory}, Cambridge, MA: EMW Publishing, 2005.

\bibitem{PEMC} Lindell, I.~V.  and A.~Sihvola, ``Perfect electromagnetic conductor", {\it J. Electro.\ Waves Appl.} Vol.~19, No.~7, 861--869, 2005.

\bibitem{DB} Lindell, I.~V.  and A.~Sihvola, ``Electromagnetic boundary condition and its realization with anisotropic metamaterial," {\it Phys.\ Rev.\ E}, Vol. 79, No.  2, 026604 (7 pages), 2009.

\bibitem{SHS1}  Kildal, P.-S., ``Definition of artificially soft and hard surfaces for electromagnetic waves," {\it Electron.\ Lett.}, Vol. 24, pp. 168--170, 1988.

\bibitem{GSH} Lindell, I. V., ``Generalized soft-and-hard surface," {\it IEEE Trans.\ Antennas Propag}, Vol. 50, No. 7, pp. 926-929, July 2002.

\bibitem{SHDB} Lindell, I.~V.  and A.~Sihvola, ``Soft-and-hard/DB boundary conditions realized by a skewon-axion medium," {\it Trans.\ IEEE Antennas Propag.}, Vol. 61, No.  2, pp. 768--774, 2013.

\bibitem{GSHDB} Lindell, I.~V.  and A.~Sihvola, ``Generalized Soft-and-hard/DB boundary," {\it ArXiv:1606.04832v1}, [physics.class-ph] 15 Jun 2016.

\bibitem{D'B'} Lindell, I.~V.  and A.~Sihvola, ``Electromagnetic boundary conditions defined in terms of normal field components," {\it IEEE Trans. Antennas Propag.}, Vol. 58, no.4, pp.1128--1135, 2010.

\bibitem{RGSH} H\"anninen, I. , I. V. Lindell and A. Sihvola, ``Realization of generalized soft-and-hard boundary," {\it Prog.\ Electromag.\ Res.}, Vol. 64, pp. 317--333, 2006.

\bibitem{Caloz} Shahvarpour, A. , T. Kodera, A. Parsa and C. Caloz, ``Arbitrary electromagnetic conductor boundaries using Faraday rotation in a grounded ferrite slab" {\it IEEE Trans.\ Microwave Theory Tech.}, Vol. 58, No. 11, pp. 2781--2793, 2010.

\bibitem{Elmaghrabi} El-Maghrabi, H. M.,  A. M. Attiya and E. A. Hashish, ``Design of a perfect electromagnetic conductor (PEMC) boundary by using periodic patches," {\it Prog.\ Electromag.\ Res.\ M}, Vol. 16, pp. 159--169, 2011.

\bibitem{Zaluski} Zaluski, D., D. Muha and S. Hrabar, ``DB boundary based on resonant metamaterial inclusions," {\it Metamaterials'2011}, Barcelona, October 2011, pp. 820--822.

\bibitem{Caloz13}  Caloz C. et al, ``Practical Realization of Perfect Electromagnetic Conductor (PEMC) Boundaries using Ferrites, Magnetless Non-reciprocal Metamaterials (MNMs) and Graphene," {\it Proc. URSI EMTS}, pp. 652--655, Hiroshima May 2013. 

\bibitem{Zaluski2014}  Zaluski, D., S. Hrabar, and D. Muha, ``Practical realization of DB metasurface," {\it Appl. Phys. Lett.}, Vol. 104, No. 234106, 2014.

\bibitem{Frezza}  Tedeschi, N., F. Frezza, and A. Sihvola, ``On the Perfectly Matched Layer and the DB boundary condition,” {\it JOSA A}, Vol. 30, pp. 1941-1946, Oct. 2013.

\bibitem{Kong08} Zhang, B., H. Chen, B.-I. Wu and J. A. Kong, ``Extraordinary surface voltage effect in the invisibility cloak with an active device inside," {\it Phys.\ Rev.\ Lett.}, Vol. 100, 063904 (4 pages), February 15, 2008.

\bibitem{Yaghjian}  Yaghjian, A. D. and S. Maci, ``Alternative derivation of electromagnetic cloaks and concentrators," {New J. Phys.}, Vol. 10, 115022 (29 pages), 2008. Corrigendum, {\it ibid}, Vol. 11, 039802 (1 page), 2009.

\bibitem{Weder} Weder, R., ``The boundary conditions for point transformed electromagnetic invisible cloaks," {\it J. Phys.\ A}, Vol. 41, 415401 (17 pages), 2008.

\bibitem{Kildal09}  Kildal, P.-S., ``Fundamental properties of canonical soft and hard surfaces, perfect magnetic conductors and the newly introduced DB surface and their relation to different practical applications included cloaking,'' {\it Proc.\ ICEAA'09}, Torino, Italy Aug. 2009, pp. 607--610.



\end{thebibliography}
\end{document}